\documentstyle[12pt]{article} \textwidth    165mm    \textheight 200mm
\begin{document}

\vspace*{0.5cm}
\centerline{PACS no.s 04.60.+n, 03.65.-w, 11.10.-z}
\centerline{Preprint IBR-TH-97-S-043, July 9, 1997}
\centerline{Contributed paper for the}
\centerline{{\it WORKSHOP ON MODERN MODIFIED THEORIES OF}}
\centerline{{\it GRAVITATION AND COSMOLOGY}}
\centerline{June 1997, Ben Gurion University, Beer Shiva, israel}

\vspace*{1cm}
\begin{center}
{\large \bf ISOMINKOWSKIAN FORMULATION OF GRAVITY}
\end{center}

\vskip 0.5cm
\centerline{{\bf Ruggero Maria Santilli}}
\centerline{Institute for Basic Research}

\centerline{P.O.Box 1577, Palm Harbor, FL 34682, U.S.A.}
\centerline{E-address ibr@gte.net; Web Site http://home1.gte.net/ibr}

\begin{abstract}
We submit the viewpoint that, perhaps, some of the controversies in
gravitation
occurred during this century are not due to
insufficiencies of Einstein's field equations,
but rather to insufficiencies in the mathematics used for their
treatment.
For this purpose we treat the same equations with the
novel, broader isomathematics and related isominkowskian geometry, and
show
an apparently final resolution in favor of existing relativities of
controversies such as: the lack of invariance of the basic
units of space and time;  lack of compatibility between
gravitational and relativistic
conservation laws; lack of
meaningful relativistic limit of gravitation; and others.
However, an apparent necessary condition for the resolution of these
controversies
is the abandonment of the notion of curvature used in this century
in favor of a conceptual and mathematical broader notion. A
number of intriguing implications
and experimental verifications are pointed out.
\end{abstract}

\vskip 0.5cm

{\bf 1. Introduction}. One of the most majestic
achievements of this century for mathematical
beauty, axiomatic  consistency and experimental verifications has been
the  {\it  special theory of  relativity}  (STR)$^{1}$. By comparison,
despite equally outstanding  achievements, the {\it  general theory of
relativity} (GTR)$^{2}$ has remained afflicted by numerous problematic
aspects at both classical and quantum levels.

The view submitted in this note is that, perhaps, some of the
controversies in gravitation
are not due to insufficiencies in current gravitational theories,
but rather to insufficiencies in their
mathematical treatment.

More specifically, we argue that the contemporary mathematics
(consisting of
conventional numbers and fields, vector and metric spaces, differential
calculus and functional analysis, etc.) has produced
an outstanding physical consistency when applied to relativistic
theories, yet the same mathematics has produced
unsettled problems when applied to gravitation.

As a concrete example, the unit I = diag. ([1, 1, 1], 1) of the
Minkowskian geometry representing in a dimensionless form
the basic units of space and time, is invariant under
the Poincar\'e symmetry, as well known.
By comparison, we have the following
\vspace*{0.5cm}

{\bf Theorem 1}$^{(3t)}$. {\it The fundamental units of space and time
are not invariant for all geometries with non-null curvature}.
\vspace*{0.5cm}

In fact,  the transition
from the  Minkowskian metric $\eta = Diag (1, 1, 1, -1)$
to a (3+1)-dimensional  Riemannian metric g(x) is characterized  by a
{\it  noncanonical}  transformation  $x\rightarrow   x' =  U\times  x,
U\times  U^{t}\not =  I$, for  which (by ignoring the dash)
$g(x) = U\times\eta\times U^{t}$. Corresponding theories of quantum
gravity
are then generally nonunitary when
formulated on conventional Hilbert spaces
over conventional complex fields. The lack of the invariance of the
basic
units then follows at both classical and operator levers form the very
definition of noncanonical and nonunitary transforms for
all gravitational theories with curvature.

Theorem 1 implies rather serious ambiguities
in the application of gravitational theories
to actual measurements, evidently because we cannot possibly
have a physically valid measure, say, of length, via a stationary meter
varying
in time. The hope that the problem is resolved by the joint change of
the
entire environment does not resolve the shortcoming because, e.g.,
the impasse remain for measures related to far away objects which, as
such,
are independent from our local environment.

We here argue that Theorem 1 is a specific manifestation of the
insufficiency
of the mathematics currently used for gravitation, because no
corresponding
shortcoming exists for the flat relativistic case.

We also argue that the shortcoming of Theorem 1 is at the foundation in
a
rather subtle way with a number of controversies in gravitation existing
in the literature. For instance,
as we shall see in this note, the achievement of a formulation of
gravity
with invariant basic units will automatically provide a novel
unambiguous operator formulation of gravity as axiomatically consistent
as relativistic
quantum mechanics. After all, no axiomatically consistent operator
theory
of gravitation should be expected without the fundamental
invariance of the basic units.

Since there is no conceivable possibility of achieving a formulation of
gravitation with invariant units based on conventional mathematics, our
basic assumption is the use of a {\it generalized mathematics} which
permits
the preservation unchanged of Einstein's field equations
and related experimental verifications while verifying the
uncompromisable conditions of invariant basic units of space and time.

The only known broader mathematics satisfying the above conditions
is the novel {\it isomathematics},
first submitted by Santilli$^{(3a)}$ back in 1978, but which achieved
sufficient operational maturity only recently in memoir$^{(3f)}$, and
which
includes: new nmumbers and fields, new vector and metric spaces, new
algebras and geometries, etc.(an outline is also available in Page 18
of Web Site [4u]).

The {\it isotopies} are nowadays referred to maps (also called liftings)
of
any given linear, local-differential, and canonical or unitary theory
into its
most general known nonlinear, nonlocal-integral and noncanonical or
nonunitary
extensions, which are nevertheless capable of reconstructing linearity,
locality and canonicity or unitarity on certain generalized
isospaces over generalized isofields.

The new geometry capable of yielding the invariance of the basic unit is
the
{\it isominkowskian geometry} first submitted by Santilli$^{(3h)}$ in
1983,
but which also reached operational maturity only recently following the
advances of the preceding memoir$^{(3f)}$.
The isominkowskian geometry was originally submitted for the most
general
possible realization of the Minkowskian axioms, but only recently has
been understood$^{(3u)}$ also to embody jointly all the machinery of the
Riemannian geometry, such as connections, covariant derivatives, etc.,
although expressed in a generalized way. In particular, the
isominkowskian
'geometry admits all possible Riemannian metrics for exterior
gravitational
models in vacuum, as well as their extensions for interior gravitational
models
with a well behaved but otherwise unrestricted dependence on the
velocities
and other interior variables.

The isominkowskian geometry therefore appears to be ideally suited for
our
objective. In fact, on one side it can preserve Einstein's
(or any other) field equations, although formulated within the context
of
a broader mathematics, while achieving the uncompromisable
invariance of the basic units of space and time.

The isotopies of classical and quantum mechanics were also submitted
by Santilli$^{(3a)}$ back in 1978, but they too reached sufficient
maturity only recently in memoir$^{(3f)}$ for the classical profile,
memoir$^{(3t)}$ for the operator profile and memoir$^{(3v)}$ for
applications
and experimental verifications.

The delay in the achievement of
operational maturity was due to the lack of invariance of preceding
studies for reasons that escaped identification for years, and which
resulted to rest where one would expect them the least,
the use of the ordinary
differential calculus under isotopies.
Once the isotopic lifting of the differential calculus
was identified in memoir$^{(3f)}$, all other problems of axiomatic
consistency
were easily solved, by reaching a generalized mathematics which yields
the invariance of the units of space and time while being
''directly universal'', that is, applicable to all well
behaved, signature preserving broadening of a given theory,
such as the Minkowskian geometry (universality), directly in the
fixed inertial frame of the observer (direct universality).

In summary, the novel isomathematics applies for the
reformulation of all possible noncanonical
or nonunitary theories while achieving the uncompromisable invariance of
the basic units of space and time.

\vskip 0.5cm

{\bf 2. Isominkowskian geometry}. The fundamental isotopy for
relativistic theories is the lifting  of the
unit of conventional theories, the unit I = diag. ([1, 1, 1], 1) of the
Minkowski space and of the Poincar\'e symmetry,
into a  well  behaved, nowhere singular,
Hermitean and   positive--definite  $4 \times 4$-dimensional
matrix $\hat {I}$    whose elements   have  an
arbitrary dependence on local quantities and, therefore, can depend on
the  space--time coordinates x and other needed variables,
$I\rightarrow \hat{I}  =  \hat{I}(x, ...)>0$.

The conventional  associative product $A\times B$  among generic
quantities  A,  B is jointly lifted   by  the {\it  inverse}  amount,
$A\times
B\rightarrow A\hat{\times}  B  = A\times  \hat{T}\times B,  \hat{I}  =
\hat{T} ^{-1}$.

Under these   assumptions $\hat{I}$ is the  (left and
right) generalized unit of  the  new theory, $\hat{I}\hat{\times}  A =
A\hat{\times}\hat{I} \equiv A,  \forall A$, in which case (only) $\hat
I$ is
called the {\it  isounit}
and $\hat{T}$  is called the  {\it isotopic element}.

For consistency,
the {\it totality} of the  original theory must be  reconstructed
to admit $\hat{I}$ as the correct (left and  right) unit. This implies
the  isotopies of    numbers,  angles,  fields,  spaces,  differential
calculus,  functional   analysis, geometries,   algebras,  symmetries,
etc. (see ref.$^{(3f)}$ for a recent account).

We now study the possible application of the above isotopies to
{\it exterior gravitation in vacuum} for which the dependence of the
isounit is
restricted to the x-dependence only. Let $M(x, \eta, R)$ be the
Minkowski space with  space--time coordinates
$x = \{x^{\mu}\} = \{r, x^{4}\}$, $x^{4} = c_{0}t$ (where $c_{0}$
is the speed of
light in vacuum), and  metric $\eta =  diag. (1,1,1,-1)$ over the reals
$R  =  R(n,+,\times)$. Let   $\Re(x,g(x),R)$ be  a $(3+1)$--dimensional
Riemannian   space with nowhere singular   and  symmetric metric $g  =
g^{t} = U\times\eta\times U^{t}$.

The regaining of the invariance of the basic units is then permitted
by assuming as basic isounit
of  the gravitational theory theory the
quantity  $\hat{I} = U\times U^{t} = \hat{I}^{t}>0$
with explicit form derivable form a Riemannian metric via the {\it
isominkowskian factorization}$^{(3o,3p)}$

\vskip 0.5cm
\begin{equation}
g(x) =   \hat{T}(x)\times\eta,  \hat{I}(x) = [T(x)]^{-1}   = U\times
U^{t}.
\label{eq:one}\end{equation}

\vskip 0.5cm
As  an example, for    the  case  of the  celebrated   Schwarzschild's
metric$^{(2d)}$,  we   have $U\times U^{t}  =  \hat{I}  =  Diag.
((1-M/r),
(1-M/r), (1-M/r), (1-M/r)^{-1})$ and similarly for other metrics. It
should
however be indicated that a better representation
of the Schwarzschild metric is that in isotropic
coordinates$^{(2f)}$ which requires a nondiagonal isounit.

        We note however that $\hat{I}$  is always positive-definite,
assured  by the locally
Minkowskian   character of Riemann. For simplicity but
without   loss  of generality, the
isounit can therefore be assumed herein as being diagonal.

An inspection of gravitational theories with {\it conventional}
Riemannian metrics $g(x) = \hat T(x)\times {\eta}$ yet referred to the
{\it generalized unit} $\hat I = \hat T^{-1}$ reveals that its axiomatic
structure is that of the {\it isotopies of the Minkowski space}$^{(3h)}$
which are characterized by the dual lifting of $\eta$ into
$\hat {\eta} = \hat T\times {\eta}$ and I into $\hat I = \hat T^{-1}$.
In fact, the {\it isotopies of Riemannian spaces}$^{(3r)}$ are
characterized by
the different dual lifting of g(x) into $\hat g = \hat T\times g$ and of
I into
$\hat I = \hat T^{-1}$.

The main difference is that in the former case the entire functional
dependence
of the metric is absorbed in the isounit, while this is not the case for
the
latter case. As we shall see, the invariance of the basic unit is
reached
in the former but not in the latter case.

The studies of gravitational theories of type (1)
must therefore be conducted within the context of the {\it isotopies of
the special relativity}, also called {\it isospecial relativity}, first
submitted by Santilli$^{(3h)}$ in 1983 and then studied in a variety
of works at both classical and operator levels (see Ref.$^{(3r,3s)}$ for
a review of the studied up to 1995 and ref.$^{(3t)}$ for studies
following the advent of the isodifferential calculus of
memoir$^{(3f)}$).

To construct    the
isospecial relativity we first  need the lifting of numbers
and  fields$^{(3g)}$. For   this    we  note  that   the    conventional
multiplicative   unit I is lifted   into  the isounit, $I  \rightarrow
U\times 1\times U^{t}    = \hat{I}$ while   the  additive unit  $0$
remains   unchanged,   $0  \rightarrow   \hat{0}  =  U\times  0\times
U^{t} =  0$.  The  numbers are  lifted   into the  so--called {\it
isonumbers},  $n \rightarrow  \hat{n}   =   U\times n\times U^{t}    =
n\times\hat{I}$ with lifting of the product $n\times m
\rightarrow \hat{n}\hat{\times}\hat{m} = \hat{n}\times\hat{T}\times
\hat{m}, \hat T = \hat I^{-1}$.

The original  field $R=R(n,+,\times  )$ is  then lifted  into the
isofield$^{(3g)}$
$\hat{R}=\hat{R}  (\hat{n},\hat{+}.\hat{\times   })$   for  which  all
operations  are isotopic. It  is easy to  see that $\hat{R}$ is locally
isomorphic  to   $R$ by   construction   and,  thus,  the  lifting  $R
\rightarrow  \hat{R}$ is   an  isotopy. Despite   its  simplicity, the
lifting is not  trivial, e.g., because the  notion of primes and other
properties of  number theory depend on the  assumed unit.  For further
aspects we refer to$^{(5r)}$ which also includes the isotopies of angles
and    functions  analysis.  Note  for   later   needs the  identity,
$\hat{n}\hat{\times }A\equiv n\times A$.

Next, we  need  the   lifting  of    the space  $M$   into  the   {\it
isominkowskian     space}$^{(3h)}$  $\hat{M}   =
\hat{M}(\hat{x},\hat{\eta
},\hat{R}$)  first proposed by Santilli in Ref.$^{3h}$
which is characterized by the {\it isocoordinates} $x
\rightarrow \hat{x} =  U\times  x\times U^{t} = x\times  \hat{I}$, and
{\it  isometric} $\eta\rightarrow \hat{\eta  }(x) = U\times \eta\times
U^{t} \equiv  g(x)$ although,   for  consistency, the latter   must be
defined on $\hat{R}$, thus having the structure $\hat{N} =
(\hat{N}_{\mu\nu}) =
\hat{\eta}\times\hat{I}=(\hat{\eta}_{\mu\nu})\times\hat{I}$.      The
conventional  interval  on  $M$ is then   lifted  into the  {\it
isointerval} on $\hat{M}$ over $\hat{R}^{4h}$

\vskip 0.5cm
\begin{eqnarray}
\lefteqn{(\hat{x}-\hat{y})^{\hat{2}}=(\hat{x}-\hat{y})^{\mu}
\hat{\times}\hat{N}_{\mu\nu}\hat{\times}(\hat{x}-\hat{y})^{\nu} =
 [(x-y)^{\mu}\times\hat{\eta}_{\mu\nu}\times
(x-y)^{\nu}]\times\hat{I}  =} \nonumber \\[0.5cm]
&=& [(x^{1}-y^{1})\times\hat{T}_{11}\times(x^{1}-y^{1})+
(x^{2}-y^{2})\times T_{22}\times(x^{2}-y^{2})+ \nonumber \\[0.5cm]
&&+(x^{3}-y^{3})\times T_{33}\times(x^{3}-y^{3})-(x^{4}-y^{4})
\times T_{44}\times(x^{4}-y^{4})\times\hat{I}.
\label{eq:two}\end{eqnarray}
\vskip 0.5cm

As one can see, the above interval coincides with the conventional
Riemannian interval by conception, except for the factor $\hat{I}$.

It is instructive to prove that the isoinvariant can also be obtained
from
the basic noncanonical transform according to the rule $\hat x^{\hat 2}
=
U\times x^2\times U^t =
[(x^t\times U^t)\times (U^{t -1}\times U^{-1})\times {\eta}
\times (U\times x)]\times (U\times U^t) =
\hat x^t\hat {\times} \hat N\hat {\times}\hat x$ where ''t'' stands for
transpose.
This construction also clarifies that the coordinates x on M are lifted
into the form $U\times x$ on ${\hat{M}}$.

It  easy to see that $\hat{M}$  is locally isomorphic  to  $M$ and the
lifting $M \rightarrow \hat{M}$ is also an isotopy. In particular, the
isospace  $\hat{M}$ is {\it isoflat},  i.e.,  it verifies the axiom of
flatness {\it in isospace over the isofields}, that is, when referred
to the generalized  unit  $\hat{I}$, otherwise $\hat{M}$ is  evidently
curved owing  to the  dependence $\hat{\eta}=\hat{\eta}(x) = g(x)$.  In
other
words,  {\it assumptions (1)  eliminate the curvature
in isospace while preserving
the Riemannian metric}. However, this is only the result of a first
inspection
of the novel isominkowskian spaces and deeper insights will soon emerge.
As we shall  these broader views on gravitation
appear to be essential
to achieve a  theory with  invariant basic units.

Note
that   $\hat{M}$ and   $\hat{R}$    have   the  {\it same     isounit}
$\hat{I}$. The conventional Minkowskian setting admitted by the
isospecial
relativity for $\hat I = i$ is therefore that in which both, the
minkowski space and ]the base
field have the same unit I = diag. ([1, 1, 1],1), which implies a
trivial
redefinition of conventional fields hereon ignored.

Studies of {\it isocontinuity  properties} on isospaces have
been  conducted by Kadeisvili$^{(4r)}$  and  those   of
the  underlying novel  {\it isotopology} by Tsagas and Sourlas$^{(4s)}$.

The {\it isominkowskian geometry}$^{(3r,3u)}$ is  the geometry of
isospaces
$\hat{M}$,  and incorporates in a   symbiotic way both the Minkowskian
and   Riemannian geometries.   In   fact, it  preserves all  geometric
properties of the  conventional {\it Minkowskian}  geometry, including
the light cone  and the  maximal causal speed $c_{o}$ (see below),
while jointly incorporating  the machinery of the Riemannian geometry
in an isotopic form. As such, it is ideally suited for our objectives.

It should be indicated that this author has studied until now
the {\it interior gravitational problem} via the {\it isotopies of the
Riemannian geometry}. The use of the {\it isominkowskian} geometry for
the characterization of {\it the exterior gravitational problem}
was briefly indicated in note$^{(3o)}$
and it is studied in more details in this work. Also, the main line of
the
isominkowskian geometry inclusive of the machinery of the Riemannian
geometry are presented in this note for the first time with
detailed study in paper$^{(3u)}$..

To outline the new geometry, one must know that, as indicated in Sect.
1,
the use
of  the  ordinary differential  calculus leads  to  inconsistencies
under
isotopies (e.g., lack    of   invariance) because  dependent    on the
assumption of the trivial unit 1 in a hidden way.  The central tool of
the isominkowskian   geometry  is therefore  the  {\it isodifferential
calculus} on $\hat{M}(\hat{x},\hat{\nu},\hat{R})$,   first  introduced
in$^{(3g)}$, which   is   characterized by  the  {\it  isodifferentials,
isoderivatives} and related properties $\hat{d}x^{\mu } = \hat{I}^{\mu
}_{\nu }\times dx^{\nu},  \hat{d}x_{\mu } = \hat{T}_{\mu }^{\nu
}\times dx_{\nu},
\hat{\partial }_{\mu }=\hat{\partial }/\hat{\partial }x^{\mu} =
\hat{T}_{\mu }^{\nu }
\times\partial /\partial x^{\nu },
\hat{\partial }^{\mu } = \hat{\partial }/\hat{\partial }x_{\mu } =
\hat{I}^{\mu }_{\nu }\times\partial /\partial x_{\nu },
\hat{\partial }x^{\mu }/\partial x^{\nu } = \delta^{\mu }_{\nu },
\hat{\partial }x_{\mu }/\hat{\partial }x^{\nu } = \hat{\eta }_
{\mu\alpha }\times \hat{\partial }x^{\alpha }/\hat{\partial }x^{\nu } =
\hat{\eta }_{\mu\nu },
\hat{\partial }x^{\mu }/\hat{\partial }x_{\nu } =
 \hat{\eta }^{\mu\alpha }\times\hat{\partial }x_{\alpha }/
\hat{\partial }x^{\nu } = \hat{\eta }^{\mu\nu }$.

Note that the original axioms must be preserved  for an isotopy. Thus, '
the isodifferential calculus is isocommutative, i.e., commutative on
$\hat M$ over $\hat R$,
$\hat{\partial }_{\alpha  }\hat{\partial}\_{\beta  } =
\hat{\partial}_{\beta  }\hat{\partial }_{\alpha  }$. However, the
same isocalculus
{\it is not}, in general, commutative in its projection on M over R.

Note    also  the    hidden  {\it   isoquotient}$^{(3g)}   A/\hat{}B   =
(A/B)\times\hat{I}$ and isoproduct  $\hat{\partial}
\hat{\times} \hat{\partial}$. Thus,
by including the isoquotient, the quantity
$\hat{\partial}\hat{\partial}$
should be more rigorously
written $\hat{\partial }\hat {\times} \hat{\partial}$. This results in
an
inessential final multiplication of the expression considered
-by $\hat {I}$ and,
as such, it will be ignored hereon for simplicity.

The entire formalism   of the {\it  Riemannian} geometry  can  then be
formulated on the  {\it iso\-min\-kow\-ski\-an} space via the
isodifferential
calculus. This aspect is  studied in details elsewhere$^{(3t)}$. We here
mention: {\it isochristoffel's symbols} $\hat {\Gamma}
_{\alpha\beta\gamma} =
\hat{\frac{1}{2}}\hat\times  (\hat{\partial }   _{\alpha    }\hat{\eta
}_{\beta\gamma } +
\hat{\partial }_{\gamma }\hat{\eta }_{\alpha\beta } -
\hat{\partial }_{\beta }\hat{\eta }_{\alpha\gamma })\times \hat{I}$,
{\it isocovariant differential}
$\hat{D}\hat{X}^{\beta} = \hat{d}\hat{X}^{\beta} +
\hat{\Gamma }_{\alpha }^{\beta }{}_{\gamma }\hat {\times}
\hat{X}^{\alpha }\hat {\times} \hat{d}\hat{x}^{\gamma }$,
{\it isocovariant       derivative}
 $\hat{X}^{\beta}_{|\hat{}\mu} =
\hat{\partial }_{\mu }\hat{X}^{\beta} +
\hat{\Gamma }_{\alpha }^{\beta }{}_{\mu } \hat {\times}
\hat{X}^{\alpha}$,
{\it isocurvature tensor}
$\hat{R}_{\alpha }^{\beta }{}_{\gamma\delta }
=\hat{\partial }_{\beta }\hat{\Gamma }_{\alpha }^{\beta }{}_{\gamma } -
\hat{\partial }_{\gamma }\hat{\Gamma }_{\alpha }^{\beta }{}_{\delta } +
\hat{\Gamma }_{p}^{\beta }{}_{\delta } \hat {\times} \hat{\Gamma
}_{\alpha }^
{p}{}_{\gamma } -
\hat{\Gamma }_{p}^{\beta }{}_{\gamma } \hat {\times} \hat{\Gamma
}_{\alpha }^
{p}{}_{\delta }$,
etc.

The verification, this time,  of the {\it Riemannian} properties
is shown by  the  fact that (under   the assumed conditions)  {\it the
isocovariant derivatives  of all isometrics  on } $\hat{M}$ {\it over}
$\hat{R}$ {\it   are   identically null},  $\hat{\eta   }_{\alpha\beta
|\hat{}\gamma} \equiv 0,\alpha , \beta , \gamma  = 1, 2,  3, 4$.  This
illustrates that    the   Ricci Lemma  also    holds  under the   {\it
Minkowskian} axioms.

A  similar occurrence holds  for all other properties, including the
five identities of the Riemannian geometry (where the firth is the
forgotten   {\it  Freud
identity}, as studied in details elsewhere$^{3u}$.

In summary, the study of the isominkowskian geometry reveals the
emergence of
the new notion of {\it isocurvature} here introduced apparently for the
first time, here referred to the redefinition of curvature via the use
of
isomathematics based on rules (1).

\vskip 0.5cm

{\bf 3. Classical unification of the special and general relativities}.
We
are now equipped to present, apparently for the first time, the
classical equations of our isominkowskian formulation
of gravity, here called
{\it  isoeinstein equations} on $\hat{M} over \hat{R}$,
which can be written

\vskip 0.5cm
\begin{equation}
\hat{G}_{\mu\nu } = \hat{R}_{\mu\nu } -
\hat{\frac{1}{2} }\hat{\times }\hat{N}_{\mu\nu }\times\hat{R} =
\hat{k} \hat{\times } \hat{\tau }_{\mu\nu },
\label{eq:three}\end{equation}

\vskip 0.5cm
where $\hat{\tau  }_{\mu\nu }$   is  the  source {\it   isotensor}  on
$\hat{M},\hat{\frac{1}{2} } = \frac{1}{2}\times\hat{I},
\hat{N}_{\mu\nu } = \hat{\eta }_{\mu\nu }\times\hat{I} =
g_{\mu\nu }\times\hat{I},   \hat{k}  = k\times \hat{I}$  and  $k$
is the  usual constant.

Despite apparent differences, it should be  indicated that
Eqs.~(3)
{\sl  coincide numerically with  Einstein's  equations} both in
isospace as well as in their projection in ordinary spaces for all
diagonal Riemannian metrics.

The preservation in isospace of the {\it numerical value} of the
conventional
field equations stems from a general property of the isotopies of
preserving
all original numerical values (see later on the preservation of the
speed of light as the maximal causal speed on $\hat M$ ). In fact,
the isoderivative $\hat{\partial}_\mu  = \hat{T}_\mu^\alpha
\times\partial_\alpha$   deviates  from   the  conventional derivative
$\partial_\mu$ by the isotopic factor $\hat{T}$.
But its numerical value  must  be referred to $\hat{I} =
\hat{T}^{-1}$, rather than $I$. This implies the preservation in
isospace of
the original value of $\partial_\mu$ and, consequently, of the original
field equations.

For the case of the projection of in ordinary spaces, the isoequations
are
reducible to the conventional equations multiplied by common isotopic
factors which, as such, are inessential and can be eliminated. In fact,
the isochristoffel's symbols~ deviate  from  the
conventional symbols by the same factor $\hat{T}$ (again, because
$\hat{\eta}
\equiv    g$), and the same happens with other terms, except for
possible redefinition of the source when needed, thus preserving again
the
conventional field equations and related experimental verifications
also in our space-time.

Note that the isominkowskian formulation of gravity permits a {\it
geometric unification  of the special  and general relativities into one
single relativity, the isospecial relativity}$^{3s}$ where for $\hat{I}
= I =    diag.(1,1,1,1)$  we have   the  special and   for $\hat{I}  =
\hat{I}(x) = U\times U^{t}$ we have the general. The invariance of the
isounit is  illustrated below.

\vskip 0.5cm

{\bf 4. Operator unification of the special and general relativities.}
We
now  indicate that the above   {\it classical} unification admit a
step--by--step {\it operator}   counterpart, called   {\it operator
isogravity} (OIG), first submitted by Santilli at the
{\it VII M. Grossmann Meeting on General Relativity} of 1993$^{(30)}$.

It should be indicated from the  outset that OIG is
structurally different than  the  conventional {\it quantum   gravity}
(QG)$^{6}$ on   numerous grounds, e.g., because  OIG  and QM have {\it
different  units, Hilbert}   spaces, and fields. In particular,  the
word
"operator" in OIG  is suggested to keep in  mind  the differences with
"quantum" mechanics (as it should also be for QG).

To identify  the explicit form of
OIG,   we  note that   the  original  {\it  noncanonical}
transform $U\times  U^{t}  = \hat{I}\neq  I $ is  mapped into a  {\it
nonunitary} transform  on a conventional Hilbert  space $\cal  H$ over
the complex field $C(c,+,\times)$.  The isounit of the operator theory
is therefore $\hat{I} = U\times U^{\dagger} =  \hat{I}^{\dagger},
\hat{T}
= (U\times U^{\dagger })^{-1} = T^{\dagger} = \hat{I}^{-1}$, where the
representation of gravity occurs  as per Eqs.  (1). Then, OIG requires
the  isotopies of the  {\it  totality}  of  {\it relativistic  quantum
mechanics} (RQM) resulting in a formulation known as {\it relativistic
hadronic mechanics} (RHM)$^{(3s,3t)}$.

Besides   the  preceding   isotopies    $R\rightarrow  \hat{R}$    and
$\hat{M}\rightarrow\hat{M}$, RHM is based on the lifting of the Hilbert
space $\cal H$  with states  $|\Psi >,|\Phi >,  ...$ and inner  product
$<\Phi |\Psi   >\in C(c,+,\times )$  into the   {\it isohilbert space}
$\hat{\cal H}^{(4t)}$ with  {\it isostetes} $|\hat{\Psi }> =
U\times|\Psi
>, |\hat{\Phi  }> =  U\times  |\Phi >,   ...,$ {\it isoinner  product}
$\hat {<\hat{\Phi }|\hat{\Psi}>} = U\times   <\Phi |\Psi    >\times
U^{\dagger   }    =  <\hat{\Phi   }|\times\hat{T}\times  |   \hat{\Psi
}>\times\hat{I}$       defined       on         the           isofield
$\hat{C}(\hat{c},\hat{+},\hat{\times     })$   with   isonormalization
$\hat {<\hat{\Psi }|\times\hat{T}\times | \hat{\Psi }>} =  {\hat I}$.

 We then have
the  {\it iso--four--momentum operator}$^{(3s,3t)}$
$p_{\mu}\hat{\times}|\hat{\Psi }>  =  -\hat
 {i}\hat {\times} \hat{\partial}_{\mu}|\hat
{\Psi }> =
-i\times \hat T_{\mu}^{\nu}\times\partial_{\nu}|\hat{\Psi }>$,   with
{\it
fundamental isocommutation rules} $[\hat{x}_{\mu }, \hat{}\hat{p}_{\nu
}] = U\times [x_{\mu }, p_{\mu }]\times U^{\dagger} =
\hat{x}_{\mu }\times\hat{T}\times\hat{p}_{\nu } -
\hat{p}_{\nu }\times\hat{T}\times\hat{x}_{\mu} =
\hat{i}\hat{\times }\hat{N}_{\mu\nu }$.
The  (nonrelativistic) {\it  isoheisenberg' equations}$^{(3b)}$ and {\it
isoschroedinger equations}$^{(3t,3u)}$ can  be  written in terms  of
the isodifferential calculus of ref.$^{(3g)}$

\vskip 0.5cm
\begin{eqnarray}
&&\hat{i}\hat{\times}\hat{d}A/\hat{d}t = i\times \hat{I}_{t}\times dA/dt
= [A,\hat{}H] =  A\times\hat{T}_{s}\times H - H\times\hat{T}_{s}\times
A, \;
\hat{I} = \hat{I}_{s}\times\tilde{I}_{t}, \nonumber \\[0.5cm]
&&\hat{i}\hat {\times}\hat {\partial}_{t}|\hat{\Psi }> =
i\times \hat{I}_{t}\times\partial_{t}|\hat{\Psi}> =  H\hat
{\times}|\hat{\Psi }>
=         \nonumber  \\[0.5cm]
& &\;\;\; = H\times\hat{T}_{s}\times|\hat{\Psi          }>             =
\hat{E}\hat{\times_{s}}|\hat{\Psi}>                                  =
(E\times\hat{I}_{s})\times\hat{T}_{s}\times|\hat{\Psi } > \equiv E\times
|\hat{\Psi }>.
\label{eq:four}\end{eqnarray}

\vskip 0.5cm
Note that the  final numbers of the theory  are conventional. We  also
have the lifting of  expectation values into the form $\hat{<}A\hat{>}
= <\hat{\Psi }|\times
\hat{T}\times A\times\hat{T}\times |\hat{\Psi } >
/< \hat{\Psi }|\times\hat{T}\times  |\hat{\Psi }>$, and the  compatible
liftings  of the  remaining   aspects of RQM$^{(3t)}$.  In   particular,
$\hat{I}$     is  the  fundamental    invariant     of    the
isotheory,
$i\hat{d}\hat{I}/\hat{d}t     =  \hat{I}\hat{\times    }H     -
H\hat{\times}\hat{I}\equiv 0$.

It  is easy   to prove that   RHM  preserves  {\it  all}  conventional
properties of RQM$^{(4t)}$. In particular: {\it isohermiticity coincides
with conventional Hermiticity},  $H^{\dagger} \equiv H^{\dagger}$  (all
quantities which  are originally   observables remain,  therefore,  so
under  isotopies); {\it the  isoeigenvalues  of isohermitean operators
are isoreal}   (thus   conventional because of    the    identity
$\hat{E}\hat{\times}|\hat{\Psi }> \equiv E\times |\hat{\Psi }>)$; {\it
RHM is form invariant under isounitary transforms} $\hat{U}\hat{\times}
\hat {U}^{\dagger } = \hat {U}^{\dagger }\hat{\times }\hat{U}  =
\hat{I}$.
In fact,
we have the invariance of the isounit $\hat{I}\rightarrow \hat{I}'=
\hat{U}\hat{\times }\hat{I}\hat{\times }hat{U}^{\dagger }\equiv
\hat{I}$,
of the  isoassociative   product    $\hat{U}\hat{\times }(A\hat{\times
}B)\hat{\times}\hat{U}^{\dagger } =
A'\hat{\times }'$;
etc;  and  the   same occurs for    all   other properties  (including
causality).  Note that nonunitary transforms on $\cal H$ can always be
identically rewritten as isounitary transforms on $\hat{\cal  H }, U =
\hat{U}
\times \hat{T}^{1/2}, U\times U^{\dagger } \equiv
\hat{U}\hat{\times }\hat{U}^{\dagger } =
\hat{U}^{\dagger }\hat{\times }\hat{U} = \hat{I}$, under which
RHM is invariant$^{(3t)}$.

It should be stressed that RHM {\it is not a new  theory, but merely a
new realization of the abstract  axioms of RQM}.  In fact, RHM and RQM
coincide at    the   abstract, realization--free    level   where   all
distinctions are  lost between $I$ and $\hat{I},  R$ and  $\hat{R}, M$
and $\hat{M},  \cal H$ and $\hat{\cal H}$, etc. Yet, RHM is inequivalent
to
RQM evidently because the two theories are related by a  nonunitary
transform. Also, RHM  is broader than
RQM,  it  recovers the latter  identically  for $\hat{I}  = I$ and can
approximate the latter as close as desired for $\hat{I} \approx I$.

On summary, the entire formulation of RHM of memoir$^{(3s,3t)}$ can be
consistently specialized for the gravitational isounit $\hat {I}(x)$
yielding the proposed OIG.

\vskip 0.5cm

{\bf 5. The Poincar\'{e}-Santilli isosymmetry.} An
important property of the isominkowskian formulation of
gravity, which is lacking  for conventional formulations, is that of
{admitting a universal, classical and operator   symmetry for all
possible
Riemannian formulations of gravitation}  first
identified by Santilli$^{(3h-3l)}$ under the name of
{\it isopoincar\'{e}  symmetry}   $\hat{P}(3.1)$, and
today called {\it Poincar\'e-Santilli isosymmetry}$^{(5,6)}$, which
results  to  be
locally isomorphic to the conventional symmetry $P(3.1)$.

The
isosymmetry $\hat P(3.1)$ is the invariance of isointerval (2) and
can   be  easly constructed via   the   {\it isotopies of Lie's
theory} first proposed by Santilli$^{(3a,3d)}$ via the lifting of
universal enveloping  algebras,  Lie   algebras,  Lie group,
transformation and  representation theories, etc., and today
called  {\it  Lie-Santilli   isotheory}$^{(5,6)}$. The latter theory
essentially  consists in the reconstruction of   all branches of Lie's
theory  for the generalized
unit $\hat{I}  = [\hat{T}]^{-1}$. Since $\hat{I} >  0$, one can see
from the inception that the Poincar\'{e}-Santilli isosymmetry is
isomorphic to the
conventional one, $\hat{P}(3.1) \approx P(3.1)$ (see ref.$^{(4t)}$ for a
recent accounts).

Note that all simple Lie algebras are known from cartan's
classification.
Therefore, the Lie-Santilli
isotheory cannit produce new Lie algebras, but only {\it new
realizations} of known Lie algebras of nonliinear, nonlocal and
nonhamiltonian type.

Moreover, a primary function of the Lie-Santilli isotheory is that of
reconstructing as exact conventional space-time and internal symmetries
when believed to be conventionally broken. In particular, one of the
primary
functions of the Poincar\'e-Santilli isosymmetry is to establish that
the
abstract axioms of the conventional Poincar\'e symmetry remnain exact
under
nonlinear, nonlocal and nonhamiltonian interactions, evidently when
properly
treated.

In this section we shall show in particular that, contrary to a rather
popular
belief, the rotational, Lorenzt and Poincar\'e symmetry do indeed remain
exact for all possible {\it gravitational} models.

The operator version of the isosymmetry $\hat P(3.1)$
is  characterized  by the
conventional generators and parameter only reformulated on isospaces
over
isofields  $X   = \{X_{k}\}  = \{M_{\mu\nu}   =
x_{\mu}p_{\nu} - x_{\nu}p,p_{\alpha }\}
\rightarrow \hat{X} = \{\hat{M}_{\mu\nu} = \hat{x}_{\mu}\times
\hat{p}_{\nu} -
\hat{x}_{\nu}\times \hat{p}_{\mu},\hat{p}_{\alpha}\}, k = 1,2,...,10,
\mu ,\nu = 1,2,3,4,$
and $w = \{w_{k}\} = \{(\theta ,v),a\}  \in R \rightarrow \hat{w}
= w\times \hat{I} \in \hat{R}(\hat{n},+,\hat{\times })$. Since the
generators of
space-time symmetries represents conventional total conservation laws,
the preservation under isotopies of conventional generators ensured {\it
ab initio}
the preservation for the isominkowskian formulation of gravity of
conventional
total conservation laws.

The isotopies
preserve the  original  connectivity properties$^{(3r)}$.  The connected
component   of  $\hat P(3.1)$    is  then   given   by
$\hat{P}_{0}(3.1)   =
S\hat{O}(3.1)\hat{\times }\hat{T}(3.1)$, where $S\hat{O}(3.1)$ is  the
{\it  connected  Lorentz-Santilli isosymmetry}
first submitted in Ref.$^{(3h)}$  and  $\hat{T}(3.1)$ is  the
group   of {\it   isotranslations}$^{3k}$. $\hat{P}_{0}(3.1)$   can be
written via the {\it  isoexponentiation} $\hat{e}^{A} = \hat{I}  +
A/1! + A\hat{\times }A/2! + . . . = (e^{A\times\hat{T}})\times\hat{I}$
characterized  by   the      {\it   isotopic
Poincar\'{e}--Birkhoff--Witt
theorem}$^{(3a,3d,5)}$ of  the    underlying isoenveloping   associative
algebra

\vskip 0.5cm
\begin{equation}
\hat{P}_{0}(3.1):\hat{A}(\hat{w})  =
\Pi_{k}\hat{e}^{i\times X\times w}               =
(\Pi_{k}e^{i\times X \times\hat{T}\times w}) \times \hat{I} =
\tilde{A}(w)\times \hat{I}.
\label{eq:five}\end{equation}

\vskip 0.5cm
Note the  appearance of the  gravitational isotopic element $\hat{T}(x)$
in the {\it exponent} of   the group structure. This illustrates   the
nontriviality   of the lifting and its   {\it nonlinear} character, as
evidently   necessary  for   any symmetry  of    gravitation. What  is
intriguing  is that the   isosymmetry $\hat P(3.1)$ recovers linearity
on
$\hat{M}$ over $\hat{R}$, a property called  {\it
isolinearity}$^{(3t)}$.

Conventional linear transforms on   $M$ {\it violate} isolinearity  on
$\hat{M}$  and  must then be    replaced with the {\it  isotransforms}
$\hat{x}' = \hat{A}(\hat{w})\hat{\times }\hat{X} =
\hat{A}(\hat{w})\times\hat{T}(x)\times\hat{x}$
which can be written   from  (5)  for computational purposes    (only)
$\hat{x}'  =   \tilde{A}(w)\times\hat{x}$.    The  preservation of   the
original      dimension   is   ensured     by     the  {\it   isotopic
Baker--Campbell--Hausdorff    Theorem}$^{(3a,3d,5,6)}$.  Structure  (5)
then
forms a connected {\it Lie--Santilli isogroup}$^{5}$ with laws
$\hat{A}(\hat{w})\hat{\times}\hat{A}(\hat{w}') =
\hat{A}(\hat{w}')\hat{\times}\hat{A}(\hat{w}) = \hat{A}(\hat{w} +
\hat{w}'),\hat{A}(\hat{w})\hat{\times}\hat{A}(-\hat{w}) = \hat{A}(0) =
\hat{I}(x) = [T(x)]^{-1}$.

As one can see,  $\hat{P}_{0}(3.1)$ is {\it  noncanonical} on $M$ over
$R$ (e.g., because it {\it does not}  preserve the conventional unit I),
but it is canonical  on $\hat{M}$  over  $\hat{R}$, a  property called
{\it isocanonicity} (because  it leaves invariant by  construction the
isounit).  This confirms  the achievement,  apparently  for the first
time, of an
operator theory of gravity verifying the fundamental invariance of its
unit. The invariance at the classical level is consequential.

One should be aware that a rather
The  use   of the isodifferential calculus on   $\hat{M}$   then  yields
the
Poincar\'e-Santilli isoalgebra $\hat{p}_{0}(3.1)^{3k}$

\vskip 0.5cm
\[
[\hat{M}_{\mu\nu     },\hat{}\hat{M}_{\alpha\beta  }]   =   i  \times
(\hat{\eta }_{\nu\alpha } \times \hat{M}_{\mu\beta } -
\hat{\eta }_{\mu\alpha } \times \hat{M}_{\nu\beta } -
\hat{\eta }_{\nu\beta } \times \hat{M}_{\mu\alpha } +
\hat{\eta }_{\mu\beta } \times \hat{M}_{\alpha\nu }),
\]
\begin{equation}
[\hat{M}_{\mu\nu }, \;\;   \hat{p}_{\alpha   }]  = i   \times
(\hat{\eta
}_{\mu\alpha } \times \hat{p}_{\nu } -
\hat{\eta }_{\nu\alpha} \times \hat{p}_{\mu}),
[\hat{p}_{\alpha },\hat{}\hat{p}_{\beta }] = 0, \;
\hat{\eta }_{\mu\nu } = g_{\mu\nu }(x),
\label{eq:six}\end{equation}

\vskip 0.5cm
\noindent where         $[A,\hat{}B]   =     A\times\hat{T}(x)\times
B     -
B\times\hat{T}(x)\times  A$   is   the  {\it   isoproduct} (originally
proposed in $^{(3b)}$),  which does  indeed satisfy  the Lie axioms   in
isospace, as one can  verify. Note the   appearance of the  Riemannian
metric $\hat{\eta  }_{\mu\nu }  =  g_{\mu\nu }(x)$,  this time, as  the
"structure      functions"   $\hat{\eta   }_{\mu\nu      }$   of    the
isoalgebra$^{3a,3d,5}$.  Note also  that the  {\it momentum components
isocommute} (while they are  notoriously non--commutative for QG). This
confirms the achievement of an isoflat representation of gravity.

The local isomorphism $\hat{p}_{0}(3.1) \approx p_{0}(3.1)$ is ensured
by the  positive--definiteness  of $\hat{T}$. In  fact,  the use of the
generators    in  the  form $\hat{M}^{\mu   }_{\nu    } = \hat{x}^{\mu
}\hat{\times }p_{\nu} -
\hat{x}^{\nu}\hat{\times}\hat{p}_{\mu }$ would yield {\it conventional}
structure constants under a {\it generalized} Lie  product, as one can
verify.  The  above   local isomorphism is   sufficient,  per se',  to
guarantee the axiomatic consistency of OIG.

The {\it isocasimir invariants}  of $\hat{p}_{0}(3.1)$ are simple
isotopic    images  of  the  conventional    ones  $C^{0}  =  \hat{I} =
[\hat{T}(x)]^{-1}, C^{(2)} = \hat{p}^{\hat{2}} =
\hat{p}_{\mu }\hat{\times }\hat{p}^{\mu } =
\hat{\eta }^{\mu\nu }\times \hat{p}_{\mu}\hat{\times}\hat{p}_{\nu},
C^{(4)} =
\hat{W}_{\mu }\hat{\times }\hat{W}^{\mu }, \hat{W}_{\mu } =
\in _{\mu\alpha\beta\pi }\hat{M}^{\alpha\beta }\hat{\times}\hat{p}^{\pi
}$.
>From     them, one can   construct     any needed {\it   gravitational
relativistic equation}, such as the {\it isodirac equation}

\vskip 0.5cm
\[
(\hat{\gamma     }^{\mu     }\hat{\times    }\hat{p}_{\mu     }    +
\hat {i} \hat {\times}\hat{m})\hat{\times
}|>  =   [\hat{\eta    }_{\mu\nu   }(x)  \times
{\hat{\gamma }}^{\mu}(x) \times \hat{T}(x)
\times\hat{p}^{\nu} - i\times m \times \hat{I}(x)] \times \hat{T}(x)
\times |> = 0,
\]
\begin{equation}
\{\hat{\gamma }^{\mu },\hat{}\hat{\gamma }^{\nu}\} =
\hat{\gamma }^{\mu}\times\hat{T}\times\hat{\gamma }^{\nu} +
\hat{\gamma }^{\nu}\times\hat{T}\times\hat{\gamma }^{\mu } =
2\times \hat{\eta}^{\mu\nu } \equiv 2\times g^{\mu\nu },
\hat{\gamma }^{\mu} = \hat{T}_{\mu\mu }^{1/2}\times
\gamma^{\mu}\times\hat{I}\;
({\rm no}\; {\rm sum}),
\label{eq:seven}\end{equation}

\vskip 0.5cm
\noindent where $\gamma^{\mu  }$  are the  conventional  gammas and
$\hat{\gamma
}^{\mu  }$ are    the  {\it isogamma   matrices}. Note  that   {\it the
anti-isocommutators  of   the    isogamma matrices  yield   (twice)  the
Riemannian    metric g(x)},  thus   confirming  the  representation of
Einstein's (or other) gravitation in   the {\it structure} of  Dirac's
equation. As an illustration, we   have the {\it Dirac--Schwarzschild
equation}   given  by    Eqs.   (7)   with    $\hat{\gamma  }_{k}    =
(1-2M/r)^{-1/2}\times\gamma_{k}\times\hat{I}$ and  $\hat{\gamma }_{4}
=   (1-2M/r)^{1/2}\times\gamma^{4}\times\hat{I}$, although, as indicated
in
Sect. 1, the representation in isotropic coordinates and a nondiagonal
isounit would be
preferable.   Similarly  one can
construct  the   isogravitational version of   all other  equations of
RQM.

These equations are not a mere mathematical curiosity because they
establish the compatibility of OIG with  experimental data in particle
physics in view   of the much  smaller contribution  of gravitational
over
electromagnetic, weak and strong contributions. Our unification of the
special and   general   relativities is,  therefore,   compatible with
experimental evidence at both classical and operator levels.

The {\it Poincar\'e-Santilli isotransforms} are given by:

{\bf 1) Isorotations}. The space components  $S\hat{O}(3)$,
called {\it isorotations}$^{(3i)}$,
can be computed from isoexponentiations  (5) with the explicit form in
the  (x,y)--plane (were we  ignore again the factorization of $\hat{I}$
for simplicity)

\vskip 0.5cm
\[
x' =
x\times\cos(\hat{T}_{11}^{\frac{1}{2}}\times\hat{T}_{22}^{\frac{1}{2}}
\times\theta_{3}) -
y\times\hat{T}_{11}^{-\frac{1}{2}}\times\hat{T}_{22}^{\frac{1}{2}}
\times
\sin(\hat{T}_{11}^{\frac{1}{2}}\times\hat{T}_{22}^{\frac{1}{2}}
\times\theta_{3}),
\]
\begin{equation}
y' =
x\times\hat{T}_{11}^{\frac{1}{2}}\times\hat{T}_{22}^{-\frac{1}{2}}\times
\sin(\hat{T}_{11}^{\frac{1}{2}}\times\hat{T}_{22}^{\frac{1}{2}}\times
\theta_{3}) + y\times\cos(\hat{T}_{11}^{\frac{1}{2}}\times\hat{T}_{22}^
{\frac{1}{2}}\times\theta_{3}),
\label{eq:e8}\end{equation}

\vskip 0.5cm
\noindent (see$^{(3s)}$   for     general  isorotations    in    all
there   Euler
angles).   Isotransforms  (8)   leave   invariant  all   ellipsoidical
deformations $x\times\hat{T}_{11}\times x + y\times\hat{T}_{22}\times y
+ z\times\hat{T}_{33}\times z = R$ of the sphere $x\times x + y\times y
+ z\times z = r$.
Such ellipsoid become perfect spheres $\hat{r}^{\hat{2}} =
(\hat{r}^{t} \times \hat{\delta } \times \hat{r}) \times \hat{I}_{s}$
in {\it isoeuclidean spaces}$^{(3h,3r)} \hat{E} (\hat{r}, \hat{\delta },
\hat{R}),
\hat{r} = \{\hat{r}^{k}\} = \{r^{k}\} \times \hat{I}_{s}, \hat{\delta }
=
\hat{T}_{s} \times \delta, \delta = diag.(1,1,1), \hat{T}_{s} =
diag.(\hat{T}_{11}, \hat{T}_{22}, \hat{T}_{33}), \hat{I}_{s} =
\hat{T}_{s}^{-1}$, called {\it isospheres}.

In fact, the deformation of the semi-axes $1_{k} \rightarrow
\hat{T}_{kk}$ while
the related units are deformed of the {\it inverse} amounts
$1_{k}\rightarrow\hat{T}_{kk}^{-1}$ preserves the perfect spheridicity
(because
the invariant in isospace is $(Length)^{2}\times (Unit)^{2})$. Note that
this
perfect sphericity in $\hat{E}$ is the geometric origin of the
isomorphism
$\hat{O}(3) \equiv O(3)$, with consequential preservation of the exact
{\it rotational} symmetry for the space--components $g(r)$ of all
possible
{\it Riemannian} metrics (becomes the isogeodesics are perfect circles).

{\bf 2) Isoboosts}. The connected Lorebntz-Santilli
isosymmetry $S\hat{O}(3.1)$ is characterized by the
isorotations and the {\it isoboosts}$^{(3h)}$ which can be written in
the
$(3,4)$--plane

\vskip 0.5cm
\begin{eqnarray}
x^{3}{}'&=& x^{3}\times\sinh(\hat{T}_{33}^{\frac{1}{2}}\times
\hat{T}_{44}^{\frac{1}{2}}
\times v) - x^{4}\times\hat{T}_{33}^{-\frac{1}{2}}\times\hat{T}_{44}^
{\frac{1}{2}}\times
\cosh(\hat{T}_{33}^{\frac{1}{2}}\times\hat{T}_{44}\times       v)     =
\nonumber \\[0.5cm]
&& = \tilde{\gamma}\times
(x^{3}-\hat{T}_{33}^{-\frac{1}{2}}\times\hat{T}_{44}^{\frac{1}{2}}\times
\hat{\beta}\times  x^{4})
\nonumber \\[0.5cm]
x^{4}{}' &=& -x^{3}\times\hat{T}_{33}^{\frac{1}{2}}\times
c_{0}^{-1}\times\hat{T}_{44}^{-\frac{1}{2}}\times\sinh
(\hat{T}_{33}^{\frac{1}{2}}\times\hat{T}_{44}\times v)  +  x^{4}\times
\cosh (\hat{T}_{33}^{\frac{1}{2}}\times\hat{T}_{44}^{\frac{1}{2}}\times
v) = \nonumber \\[0.5cm]
&=&\tilde{\gamma }\times (x^{4}-\hat{T}_{33}^{\frac{1}{2}}
\times\hat{T}_{44}^{-\frac{1}{2}}\times\tilde{\beta}\times x^{3})
\nonumber \\[0.5cm]
\tilde{\beta^2} &=& v_{k}\times\hat {T}_{kk}\times v_{k}
/c_{0}\times \hat{T}_{44}\times c_{0} ,\;\;\;
\tilde{\gamma} = (1-\tilde{\beta}^{2})^{-\frac{1}{2}}
.
\label{eq:9}\end{eqnarray}

\vskip 0.5cm
Note that the above isotransforms are formally similar to the Lorentz
transforms,
as expected from their isotopic character. Isotransforms (9)
characterize the
{\it light isocone}$^{((3s))}$, i.e., the perfect cone in isospace
$\hat{M}$. In a
way similar to the isosphere, we have  the deformation of the light cone
axes
$1_{\mu} \rightarrow \hat{T}_{\mu\mu }$ while the corresponding units
are
deformed of the {\it inverse} amount $1_{\mu} \rightarrow
\hat{T}_{\mu\mu}^{-1}$,
thus preserving the perfect cone in isospace.

In particular, the isolight cone also has the conventional
characteristic
angle, as a
necessary condition for an isotopy (the proof of the latter property
requires
the use of isotrigonometric and isohyperbolic functions). Thus,
{\it the maximal
causal speed in isominkowski space is the conventional speed in vacuum}
$c_{0}$. The
identity of the light cone and isocones is the geometric origin of the
isomorphism $S\hat{O}(3.1) \approx SO(3.1)$ and, thus, of the exact
validity of
the {\it Lorentz} symmetry for all possible {\it Riemannian} metrics
$g(x)$.

{\bf 3) Isotranslations}. The {\it isotopies of translations} can be
written
\vskip 0.5cm
\begin{equation}
x' = (\hat{e}^{i\times\hat{p}\times a})\hat{\times}\hat{x} =
[x + a\times A(x)]\times\hat{I},
\hat{p}' = (\hat{e}^{i\times\hat{p}\times a})
\hat{\times}\hat{p} = \hat{p},\\ [0.5cm]
A_{\mu} = \hat{T}_{\mu\mu }^{1/2} + a^{\alpha }
\times[\hat{T}_{\mu\mu }^{1/2},
\hat{}\hat{p}_{\alpha }]/1! + .....
\label{eq:10}\end{equation}

\vskip 0.5cm
\noindent and they are also nonlinear, as expected.

{\bf 4) Isoselftransforms}. Intriguingly, the isotopies identify
one additional symmetry which is absent in the conventional case. It is
here
called {\it isoselfscalar invariance} and it is given by
\vskip 0.5cm
\begin{equation}
\hat{I} \rightarrow\ \hat{I}' = n^{2}\times\hat{I},
\eta \rightarrow\ \hat{\eta} = n^2\times \eta,
\label{eq:e11}\end{equation}

\vskip 0.5cm
\noindent where n is
an 11-th parameter, under which the interval
remains invariant, $\hat{x}^{\hat{2}} =
(x^{\mu }\times\hat{T}_{\mu}^{\alpha }\times\eta_{\alpha\nu }\times
x^{\nu})\times\hat{I} \equiv
[x_{\mu}\times (n^{-2}\times\hat{T}_{\mu}^{\alpha})\times\eta_{\alpha\nu
}\times x^{\nu}]\times(n^{2}\times\hat{I})$.

Note that, even though $n^{2}$ is factorizable, the corresponding
isosymmetry
is not trivial, e.g., because $n^{2}$ enters into the {\it argument} of
the
isolorentz transforms $(9)$. Note also that the isominkowskian
representation
of gravity is permitted precisely by the latter isoinvariance.
In fact, isoinvariance (11) holds
also for the conventional Poincar\'e symmetry, by introducing the
generalized unit
at the foundation of the isominkowskian gravity.

The same symmetry also holds for the isoinner
product (whenever $n$ does not depend on the integration variable),
$<\hat{\Phi}|\times\hat{T}\times |\hat{\Psi}>\times\hat{I} \equiv
<\hat{\Phi}|\times (n^{-2}\times\hat{T})\times
|\hat{\Psi}>\times(n^{2}\times{I})$.
Note finally that the latter symmetries have remained undetected
throughout this
century because they required the prior discovery of {\it new numbers},
those
with an arbitrary unit$^{3g}$.

{\bf 5) Isoinversions}. The  {\it     isodiscrete
transforms}$^{(3i)}$
are
\vskip 0.5cm
\begin{equation}
\pi\times x = (-r,x^{4}), \hat{\tau }\hat{\times }x = \tau\times x =
(r,-x^{4}),
\hat{\pi } = \pi\times\hat{I},  \hat{\tau } = \tau\times\hat{I},
\label{eq:twelve}\end{equation}

\vskip 0.5cm
\noindent where  $\pi$, $\tau$ are the   conventional inversion
operators. Despite
their simplicity, the physical implications of isoinversions
are nontrivial because of
{\it the possibility    of  reconstructing as  exact
discrete  symmetries when believed  to be  broken}, which is studied by
embedding all symmetry breaking
terms in the isounit$^{(3s)}$

The {\it general Poincar\'e-Santilli isosymmetry} is usually
defined as the 11-dimensional
set of {\it isorotations, isoboosts, isotranslations, isoinversions,
isoselftransforms and isoinversions}. The {\it restricted
Poincar\'{e}-Santilli isosymmetry} is the general isosymmetry in which
the isounit is averaged into constants.

\vskip 0.5cm

{\bf 6. Inclusion of interior gravitation.} The attentive
reader may have noted that the isotopies leave unrestricted the
functional dependence of the isometric. Its sole
dependence on the coordinates is therefore
a {\it restriction} which has been used so far for a
representation of {\it exterior
gravitation in vacuum}.

In the general case we have isometrics with an unrestricted
functional dependence,
$\hat{\eta} = \hat{T}(x,v, \Psi, \partial{\Psi}, ...)\times\eta ,
\hat{T} > 0,
v = dx/dt$,
which, as such, can represent {\it interior gravitation problems} with a
well behaved but otherwise
{\it unrestricted nonlinearity in the velocities, wave functions and
their
derivatives}, as expected in realistic interior models, e.g., of neutron
stars,
quasars, black holes and all that.

Note also that the isometric can also contain
{\it nonlocal--integral terms}, e.g., representing
wave-overlappings$^{(3s)}$.
Nevertheless, the theory verifies the condition of locality in isospace,
called
{\it isolocality}, because its topology is everywhere local except at
the
unit$^{(3t,4r)}$.

Note that the addition of interior gravitational problems occurs without
altering the
axioms of the exterior problem in vacuum, yet gaining an
arbitrary functional dependence for more realistic treatments of
interior conditions.
This evidently permits a geometric unification of exterior and interior
gravitational
problems which are solely differentiated by the functional dependence of
the isounits.

A first illustration of the extension of the exterior axioms to
realistic interior conditions is offered by the isoselfscalar transforms
(11) which
permit the {\it representation of electromagnetic waves propagating
within
physical media with local varying speed} $c=c_{0}/n$.

This allows the construction,
apparently for the first time, of {\it Schwarzschild's and other
gravitational
models within atmospheres and chromospheres with a locally varying
speed of light}. Applications to specific cases, such as the study of
gravitational
horizons via the light isocone, are then expected to
permit refinements of current studies evidently
due to deviations from the value in vacuum of the speed of light in the
hyperdense chromospheres outside gravitational horizons.

Note that the general Poincart\'e-Santilli isosymmetry is used for the
{\it local} speed of light within the interior of
a given atmosphere or chromosphere, while the restricted
isosymmetry is used when the {\it average} speed of light is needed.

As an example, the characterization of the speed of electromagnetic
waves traveling
within a planetary atmosphere or chromosphere requires the general
Poincar\'e-Santilli
isosymmetry because changing with the density, temperature, etc. On the
contrary,
the characterization of the average speed of electromagnetic waves
propagating
through an entire given atmosphere or chroomosphere requires the use of
the
restricted isosymmetry.

We finally note that the realization of the isotopic element
$\hat T_{\mu \mu} = n^{-2}, \mu = 1, 2, 3, 4$ is a particular case of
the broader
realization $\hat T_{\mu \mu} = n_{\mu}^{-2}$ in which the index of
refraction
is $n_4$ and the $n_k's$ provide its "space-time symmetrization". The
latter
realization is particularly suited for the direct geometrization
of the {\it anisotropy} of astrophysical atmospheres or chromospheres
caused by intrinsic angular momenta, as well as their {\it
inhomogeneity}
caused by the radial change of density and other characteristics.
\vskip 0.5cm

 {\bf 7. Direct universality of the Poincar\'e-Santilli isosymmetry for
exterior and interior gravitations.} The results of this note imply the
following:

\vskip 0.5cm
{\bf Theorem 2.} {\sl The 11-dimensional, general, Poincar\'{e}-Santilli
isosymmetry on isominkowski spaces
over real isofields with well behaved, positive-definite isounits
is the
largest possible isolinear, isolocal and isocanonical invariance
of isoseparation $(2)$ (universality) in the fixed x-frame of the
experimenter (direct universality).}

\vskip 0.3cm
The verification of the invariant under the Poincar\'{e}-Santilli
isotransforms of all
possible separation $(2)$ is instructive. The maximal character of the
isosymmetry can be proved as in the conventional case. Note that for any
arbitrarily given (diagonal) Riemannian metric $g(x)$ (such as
Schwarzschild,
Krasner, etc,) {\it there is nothing to compute} because one merely {\it
plots}
the $\hat{T}_{\mu\mu }$ terms of the decomposition
$g_{\mu\mu } = \hat{T}_{\mu\mu }\times\eta_{\mu\mu }$
(no sum) in the above given isotransforms. The invariance of the
separation
$x^{t}\times g\times x$
is then ensured.

The $(2+2)$--de Sitter or other cases can be derived from the
theorem via mere changes of signature or dimension of the isounit. The
extension to
positive-definite yet nondiagonal isounit is trivial and will be implied
hereon.

Note finally that isosymmetry $\hat P(3.1)$ cannot be even defined,
let alone constructed in conventional Riemannian spaces and
all their possible isotopies, thus rendering the isominkowskian
formulation of gravity rather unique for our purposes.

\vskip 0.5cm

{\bf 8. Resolution of some of the controversies in gravitation.} In
summary, in this note we have presented,
apparently for the first time, {\it a geometric unification of the
special and general relativities for both classical and operator
profiles,
as well as for both exterior and interior problems.} The results
are centrally dependent on the use of the {\it isominkowskian} geometry
introduced in this note and Ref.$^{(3u)}$,
rather than the use of the {\it isoriemannian} form as studied in
Ref.$^{(3s)}$.

The above unifications are
centrally dependent on the achievement of a universal symmetry for
gravitation
which, by conception and construction, is locally isomorphic to the
Poincar\'{e}
symmetry of the special relativity. This eliminates the
historical difference between the special and general
relativities whereby the former admits a universal symmetry, while the
latter
does not$^{(1,2)}$. Note the {\it necessity} of the
representation of gravity in {\it isominkowski} space for the very
formulation
of the above unifications.

These results have a number of implications. First, they
allow to illustrate the viewpoint expressed in Sect. 1 to
the effect that some of controversies in gravitation
may well be due to
insufficiencies in the used mathematics.

The first illustration is given by the physical shortcoming of
conventional
formulation of gravitation of being without invariant basic units of
space and time (Theorem 1). This shortcoming can now be rigorous
verified
via Theorem 2.
In fact, it is easy to see that,
when formulated on conventional spaces over conventional fields,
the isosymmetry $\hat P(3.1)$ {\it does not}
leave invariant conventional units.

Theorem 2 also allows to resolve the
shortcoming. In fact, the space-time isounit is indeed invariant under
the isosymmetry $\hat P(3.1)$ by conception and construction.
Moreover, it has ben proved in the adjoint analytic study [3x] that the
isosymmetry $\hat P(3.1)$ also leaves invariant the {\it conventional}
unit
I = diag. ([1, 1, 1], 1) when interpreted as isocanonical transforms on
isospaces over isofields. This is an expected consequence of the
mechanism
of isotopies according to which the joint lifting of a metric while the
basic unit is lifted by the inverse amount preserves all original
properties.

Theorem 2 also permits the resolution of the
controversy whether the total conservation laws of general relativity
are compatible with those of the special relativity
via a mere {\it visual examination}.

Recall that the generators of all space-time symmetries characterize
total
conserved quantities. The compatibility of the total conservation laws
of the general and special relativities is therefore established by the
visual observation that {\it the generators of the conventional and
isotopic
Poincare'
symmetries coincide}. In fact, only the {\it mathematical operations} on
them
are changed in the transition from the relativistic to the gravitational
case.

Yet another controversy which appears to be resolved by our
isominkowskian treatment of gravity is the apparent lack of a meaningful
relativist limit in conventional gravitational theories. In fact, such a
limit is
now clearly and unequivocally established by $\hat I \rightarrow I$
under which
the special relativity is recovered identically in all its aspects.

The isominkowskian treatment of gravity also permits a resolution of
some of
the limitations of conventional gravitational models, such as their
insufficiency to provide an effective representation of {\it interior}
gravitational problems. In fact, conventional formulations of gravity
admit
only a limited dependence on the velocities, while being strictly
local-differential and derivable from a first-order Lagrangian
(variationally
self-adjoint$^{(3d)}$). These
characteristics are evidently exact for exterior problems in vacuum.

By comparison, interior gravitational problems, such as all forms of
gravitational collapse, are constituted by extended and hyperdense
hadrons
in conditions of total mutual penetration in large numbers into small
regions of space. It is well known that these conditions imply effects
which
are {\it arbitrarily nonlinear in the
velocities as well as in the wavefunctions, nonlocal-integral on
various quantities and variationally nonselfadjoint}$^{(3d,3e)}$, (i.e.
not
representable via first-order Lagrangians). It is evident that the
latter
conditions are beyond any scientific expectation of quantitative
treatment
via conventional gravitational theories.

The isominkowskian formulation of gravity resolve this limitation too
and
shows that it is equally due to insufficiencies in the underlying
mathematics.
In fact, isogravitation extends the applicability of Einstein's axioms
to a form which is "directly universal" for
exterior and interior gravitations.

As indicated earlier, this extension is due to the fact
that the functional dependence of the metric in Riemannian treatments
is restricted
to the sole dependence on the local coordinates, g = g(x), while under
isotopies the same dependence becomes unrestricted,
$g = g(x, v, \phi, \partial{\psi}, ...)$ {\it without altering the
original
geometric axioms}. This results in {\it geometric unification of
exterior and interior problems}, despite their sizable structural
differences of
topological, analytic and other characters. The latter unification
was studied in details in ref.$^{(3s)}$  under
the {\it isoriemannian} geometry and it is studied with the
{\it isominkowskian} geometry in this note for the reasons indicated
earlier.

Yet another controversy which appears to be resolved by the
isominkowskian
formulation of gravity is the achievement of an axiomatically consistent
operator
version of gravity, that is with: invariance of the basic units;
preservation of the original Hermiticity at all times; uniqueness and
invariance of the numerical predictions; consistent PCT and other
theorems; etc.

Even though far from being a complete theory, our OIG
does indeed offer realistic hopes of achieving such an axiomatically
consistent operator form of gravity, as expected from the validity of
the
{\it conventional} axioms of RQM.

The resolution of other controversies cannot be studied in this
introductory note
and are contemplated for study in subsequent works.
\vskip 0.5cm

{\bf Concluding Remarks}. By keeping in mind that Einstein's field
equations are
preserved unchanged by conception, an important issue is whether the
isominkowskian reformulation of gravity coincides with conventional
gravity on
physical grounds or it predicts novel physical features.

It is easy to see that a number of new features are indeed predicted.
To begin, {\it the
isominkowskian formulation of gravity predicts that the maximal causal
speed in
our space-time is a local quantity which can be arbitrarily smaller
or bigger 'than the speed of light in vacuum.}

In fact, except for being well behaved (and non-null),
the parameter n of isoselftransforms (11) remains unrestricted by the
isotopies. Therefore, we have $n=1$ in vacuum for which $c = c_o/n =
c_o$,
but otherwise we can have $n > 1 (c < c_o)$ or $n < 1 (c >c_o)$.
As a result, {\it the Poincar\'{e}-Santilli
isosymmetry is a natural invariance for arbitrary
causal speeds, whether equal, smaller or bigger than the speed of light
in vacuum}.

The case $c < c_o$ is known since Lorentz's$^{(7a)}$
who was the first to investigate the lack of applicability of his
celebrated transforms for electromagnetic waves propagating in our
atmosphere or other transparent material media (see also the related
quotation by
Pauli$^{(7b)}$).

The case $c > c_o$ has been predicted since some time {\it in interior
problems only}, but experimentally detected only recently, e.g., for the
speed
of photons traveling in certain guides$^{(8a,8b)}$ or for the speed of
matter in
astrophysical explosions$^{(8c-8e)}$. The recent Ref.$^{(8f)}$ has
identified
solutions of {\it conventional}
relativistic equations with {\it arbitrary speeds in vacuum} of
which $\hat{P}(3.1)$ is evidently the natural invariance). If confirmed,
these
waves would be the first case of speeds $c > c_o$ in
{\it exterior conditions in vacuum.}

It should be noted that,
despite the local variation of $c$, the maximal causal speed on
$\hat{M}$ over
$\hat{R}$ remain $c_{0}$, again, because the change $c \rightarrow
c_{0}/n$ is
compensated by an inverse change of the unit $1\rightarrow n$.
By recalling that the STR is
evidently inapplicable (and not "violated") for locally
varying causal speeds, we
can therefore say that {\it the isotopies render the STR universally
applicable
for relativistic and gravitational, classical and operator, as well as
exterior or interior problems, under local speeds of electromagnetic
waves}.

Yet other novel predictions are related to gravitational singularities.
In fact, we have the following property of self-0evident proof.
\vskip 0.5cm

{\bf Theorem 3.} {\it Gravitational singularities (horizons) are the
zeros of
the space (time) component of the isounit.}
\vskip 0.5cm

The above novel interpretation of gravitational singularities and
horizons
is trivially equivalent to the conventional one for the
{\it exterior} case in vacuum.
However, gravitational collapse is a typical {\it interior} case for
which the
isotopic representation becomes nontrivial, e.g., because it includes
the nonlinear, nonlocal and noncanonical effects indicated earlier. The
isominkowskian formulation of gravity therefore implies a re-examination
of gravitational singularities on both mathematical and physical grounds
which will
be done elsewhere.

Note that the zeros of the isounit have been excluded from Theorem 2
because of
their yet unknown topological structure.

Another important implication of the isominkowskian formulation of
gravity is that
it offers realistic possibilities for an axiomatically consistent
inclusion of gravitation
in grand unified gauge theories, as studied in the recent
contribution$^{(3y)}$
to the {\it VIII M. Grossmann Meeting on General Relativity}. By
comparison,
no such inclusion is possible for the Riemannian treatment of
gravitation.

Intriguingly, the {\it Isotopic Grand Unification} is permitted
precisely by
the elimination of the conventional notion of curvature, as an evident
necessary
condition to bring gravitation into a form axiomatically compatible with
electroweak interactions.

We should also indicate the novel prediction of the {\it isodoppler
shift}$^{(3t)}$, namely, a shift due to the {\it inhomogeneity and
anisotropy} of the medium
in which electromagnetic waves propagate, which is suitable for
experimental
verifications$^{(3s,5b)}$.

But perhaps the most intriguing novel feature of the isominkowskian
formulation of
gravity is that of introducing {\it a novel notion of space-time},
where the novelty rests in the {\it units} of space-time itself.
For instance, we have a novel local notion of time,
as illustrated by the dependence of its unit from the gravitational
field in the
isotopic reformulation of the Schwarzschild metric, $\hat I_t = (1 -
2M/r)$.

As one can see, time is predicted to have novel different flows for
different
gravitational fields, according to a behaviour which
is different than that predicted by conventional
gravitational theories. Space has a behaviour which is the inverse of
that of time.
Conventional space-time is recovered in empty space for M = 0 (or for
infinite distances from a gravitational field).

This novel notion of space-time admits additional
intriguing and far reaching implication.
For instance, by recalling that the structure of the isotopic invariant
is
$[Length]^2\times [Unit]^2 = Inv.$ we have a new form of {\it geometric
propulsion}
called {\it isolocomotion}$^{(3r)}$ in which distances are covered by
their
geometric reduction due to the local increase of the energy, rather than
via
Newtonian displacement. In fact, the space-isounit increases with the
local energy,
thus impllying a decrease of the distance.

Similarly, we have a {\it novel cosmology} called {\it
isocosmology}$^{(3s)}$
which is the first on scientific record being {\it characterized by a
universal symmetry}, the Poincar\'e-Santilli isosymmetry $\hat P(3.1)$,
the first to
admit a realistic representation of interior gravitational problem and
the
first to admit arbitrary local values of the maximal causal speed, with
a
number of consequences, such as the elimination of the need for the
"missing mass" in the universe (see ref. [3s] for brevity).

It should be also indicated that the isotopies with basic lifting
$I \rightarrow \hat{I}(x,\Psi ,...) = \hat{I^{\dagger}}$
constitute only the first step of a chain of generalized
methods$^{(3f)}$. The
second class is given by the {\it genotopies}$^{(3a)}$ in which the
isounit is no
longer Hermitean. This broader class geometrizes in a natural way the
irreversibility of interior gravitational problems
and it has been used, e.g., for the black hole model of
ref.$^{(4d)}$. A third class of methods is given by the (multi--valued)
{\it hyperstructures}$^{(3f)}$, in which the generalized unit is
constituted by a
{\it set} of non-Hermitean quantities. The latter class
appears to be significant for quantitative studies of biological
structures with their
typical irreversibility and variation of physical characteristics
for which the conventional RQM is manifestly inapplicable due to its
reversibility as well as sole characterization of conservation laws.

Also, the isotopies, genotopies and hyperstructures admit
antiautomorphic
images, called {\it isodualities}$^{(3r,3s)}$, and characterized by the
map
$\hat{I} \rightarrow \hat{I}^{d} = -\hat{I}^{\dagger}$ which are
currently
under study for antimatter$^{(3q)}$. In this case the energy--momentum
tensor of
antimatter becomes {\it negative--definite}, thus removing a problem of
compatibility between the current representations of antimatter in
classical and
particle physics. The gravitational treatment of antimatter via the
isodualities of the isominkowskian geometry is studied
elsewhere$^{(3u)}$ and it
is another necessary condition for the axiomatically consistent
inclusion of gravity
in unified gauge theories of electroweak interactions$^{(3y)}$.

On historical grounds, we note that, as studied in detail in
memoir$^{(3t)}$
for the general case of RHM, our OIG can be interpreted as a
{\it nonunitary completion} of RQM considerably along the historical
$E-P-R$
argument$^{(9a)}$ for which von Neumann's theorem$^{(9b)}$
and Bell's inequalities$^{(9c)}$
do not apply evidently because of their nonunitary structure.

Moreover, from the
abstract identity of the right modular associative action
$H\times |\Psi >$ and its isotopic image
$\hat{H}\hat{\times }|\hat{\Psi }>$, one can see that the
isoeigenvalue equation
$\hat{H}\hat{\times }|\hat{\Psi }> = E_{\hat{T}}\times |\hat{\Psi}>$
characterizes an explicit and concrete {\it operator realization of the
"hidden
variable"} $\lambda = \lambda (x,...)\equiv \hat{T}$. Our isotopic
formulation
of gravity can therefore be interpreted as a realization of the theory
of hidden
variables. After all, the "hidden" character of gravitation in our
theory is
illustrated by the recovering of the conventional unit under the
isoexpectation
value $\hat{<}\hat{I}\hat{>} = I$.

In conclusion, the viewpoint we have attempted to convey in this note is
that
an alternative formulation of gravity always existed.
It did creep in un--noticed until now because embedded where nobody
looked for,
in the {\it unit} of the classical and quantum version of the special
relativity.

\vskip 0.5cm

{\bf Acknowledgments.} The author would like to
thank for invaluable comments the participants
to: the {\sl VII Marcel Grossmann Meeting  on General Relativity} held
at
Stanford university in July   1994; the {\sl International  Workshops}
held at the Istituto   per la Ricerca di  Base  in Molise,   Italy, on
August  1995 and  May 1996;  and the  {\sl  International Workshop  on
Physical Interpretation of Relativity  Theories} held at the  Imperial
College, London, on September 1996; the {\it VIII
Marcel Grossmann Meeting  on General Relativity} held at the Hebrew
University
in Jerusalem on June 1997 and the
{\it Workshop on Modern Modified Theories of Gravitation
and Cosmology} held at the Ben Gurion University, Israel, on June 1997.
The author would like also to express his sincere appreciation
to the Editors of
{\it Rendiconti Circolo Matematico Palermo},
{\it Foundations of Physics} and {\it Mathematical Methods in Applied
Sciences},
for invaluable, penetrating and
constructive critical comments in the editorial processing of the
respective
memoirs$^{(3f,3t,4q)}$, without which this paper could not have seen
the light.  Thanks are finally
due to L. P. Horwitz of Tel Aviv University and
E. I. Guendelman of Ben Gurion University, Israel, for
penetrating critical comments and suggestions.

\end{document}